%% file: ms.tex
\documentclass[11pt]{article}

\usepackage{times}
\usepackage{authblk}
\usepackage[pdftex]{graphicx}
\usepackage{xcolor}
\usepackage{color}
\usepackage{colortbl}
\usepackage{xspace}
\usepackage{amsmath,amscd,amssymb}
\usepackage{subfigure}
\usepackage{boxedminipage}
\usepackage{comment}
\usepackage{url}
\usepackage{enumitem}
\usepackage{wrapfig}
\usepackage{booktabs} 
\usepackage{mathrsfs}
\usepackage{balance}
\usepackage{algorithmic}
\usepackage[algoruled]{algorithm2e}
\usepackage{cite,alltt}
\usepackage{float}
\usepackage[rflt]{floatflt}
\usepackage{array}
\usepackage[labelfont=bf]{caption}

\newtheorem{definition}{Definition}

\newcommand{\ie}{{i.e.,}\xspace}
\newcommand{\eg}{{e.g.,}\xspace}

\newcommand{\datasets}{data sets\xspace}

\newcommand{\myparagraph}[1]{\vspace{0.1cm}\noindent \textbf{#1}}
\newcommand{\myparagraphem}[1]{\vspace{0.05cm}\noindent \emph{#1}}
\newcommand{\itemspace}{\vspace{-.1cm}}
\newcommand{\rasterjoin}{RasterJoin\xspace}

{\begin{list}{$\bullet$}{%
        \setlength{\labelsep}{2pt}\setlength{\leftmargin}{0pt}%
        \setlength{\labelwidth}{0pt}%
        \setlength{\listparindent}{0pt}}}
{\end{list}}
\newcommand{\denselist}{\itemsep 0pt\parsep=1pt\partopsep 0pt}

\def\R{\mathbb{R}}

\def\C{\mathbb{C}}

\def\opG{\mathscr{G}}
\def\opV{\mathscr{V}}
\def\opM{\mathscr{M}}
\def\opB{\mathscr{B}}
\def\opD{\mathscr{D}}

\AtBeginDocument{%
	\providecommand\BibTeX{{%
			\normalfont B\kern-0.5em{\scshape i\kern-0.25em b}\kern-0.8em\TeX}}}
		
\begin{document}

\title{A GPU-friendly Geometric Data Model and Algebra for Spatial Queries: Extended Version}

\author[]{Harish Doraiswamy}
\author[]{Juliana Freire}
\affil[]{New York University}
\affil[]{\textit{harishd@nyu.edu} \hspace{1cm} \textit{juliana.freire@nyu.edu}}
\date{}

\maketitle
\input{abstract}

\input{intro-new}
\input{representation}
\input{operators}
\input{queries}
\input{implementation}
\input{exp}
\input{discussion}
\input{rel-work}

\input{conclusion}

\subsection*{Acknowledgements} This work was partially supported by the
DARPA D3M program and the NYU Moore Sloan Data Science Environment.

%
\bibliographystyle{abbrv}
\bibliography{paper}  

\end{document}

%% file: abstract.tex
\begin{abstract}
  The availability of low cost sensors has led to an unprecedented
  growth in the volume of spatial data. However, the time required to
  evaluate even simple spatial queries over large data sets greatly
  hampers our ability to interactively explore these data sets and extract
  actionable insights.
  Graphics Processing Units~(GPUs) are increasingly being used to
  speedup spatial queries. However, existing
  GPU-based solutions have two important drawbacks: they are
  often tightly coupled to the specific query types they target,
  making it hard to adapt them for other queries; and
  since their design is based on CPU-based approaches, it
  can be difficult to effectively utilize all the benefits provided by
  the GPU.
  As a first step towards making GPU spatial query processing
  mainstream, we propose a new model that represents spatial data as
  geometric objects and define an algebra consisting of GPU-friendly
  composable operators that operate over these objects.  We
  demonstrate the expressiveness of the proposed algebra by
  formulating standard spatial queries as algebraic expressions.
  We also present a proof-of-concept prototype that supports a subset
  of the operators and show that it is at least two orders of
  magnitude faster than a CPU-based implementation. This performance
  gain is obtained both using a discrete Nvidia mobile GPU and the
  less powerful integrated GPUs common in commodity laptops.
\end{abstract}

%% file: intro-new.tex
\section{Introduction}
\label{sec:intro}

The availability of low cost sensors 
such as GPS in vehicles,  mobile and IoT devices have led 
to an unprecedented growth in the volume of spatial data. 
Extracting insights from these data sets requires the ability to
effectively and efficiently handle a variety of queries.

The most common approach to support spatial queries is through the use of
spatial extensions that are available in existing relational database
systems (\eg the PostGIS extension for PostgreSQL~\cite{postgis}, Oracle Spatial~\cite{oracle-spatial},
DB2 Spatial Extender~\cite{db2-spatial}, SQL Server Spatial~\cite{sql-spatial}).
Popular geographic information system~(GIS) software typically use
these systems to process spatial queries~\cite{arcgis, grass, qgis}
and some also provide their own database backend.
Using these state-of-the-art systems, the response times to even
simple spatial queries over large data sets can runs into several
minutes (or more), hampering the ability to perform interactive
analytics over these data~\cite{liu-heer@tvcg2014,fekete12}.
While faster response times can be attained by powerful
clusters~\cite{Eldawy2016,pandey@vldb2018}, such an option, due to its
costs and complexity, is often out of reach for many analysts.

Recent technological advances have made Graphics Processing
Units~(GPUs) a cost-effective alternative to provide high computing
power. Since GPUs are widely available, even in commodity laptops,
effective GPU-based solutions have the potential to democratize
large-scale spatial analytics.
Not surprisingly, several approaches have been proposed that use GPUs to speed up
spatial queries~(\eg
\cite{STIG2016,Zhang2012a,Zhang2012b,Zhang2012d,Bustos2006}).
However, these implementations typically follow the
traditional approaches that were designed primarily for the CPU, and
simply porting the algorithms may lead to an ineffective use of GPU
capabilities.
For example, a spatial aggregation query that aggregates
input points
across different polygonal regions would typically be implemented as 
a spatial join of the points and polygons followed by the aggregation of the join results.
On the other hand, the recently proposed \rasterjoin~\cite{rasterjoin}
represented a departure from traditional strategies: by modeling
spatial aggregation queries using GPU-specific operations, it attained
significant speedups even over the traditional query plan executed on
a GPU, suggesting that GPU-specific strategies can lead to substantial
performance gains.

\begin{figure}[t]
	\centering
	\includegraphics[width=\linewidth]{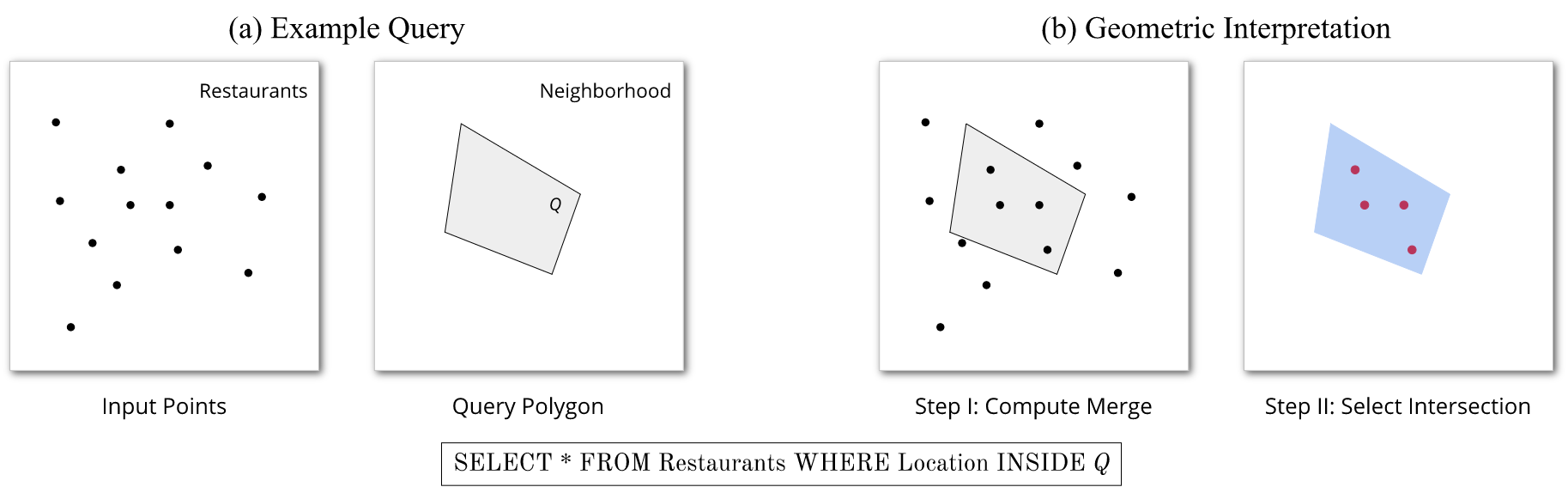}
	\vspace{-0.5cm}
	\caption{Reformulation of the spatial selection query.}
	\label{fig:example}
	\vspace{-0.25cm}
\end{figure}

Another drawback of traditional GPU-based approaches is that
they require different implementations for each query
class. Consequently, it is hard to re-use and/or extend them for
other similar queries.
As an example, consider the simple spatial selection
query illustrated in Figure~\ref{fig:example}(a).
Given a spatial data set consisting of a collection of points (say
restaurants)  and their locations,
this query identifies all points that are contained within
the specified query polygon (\eg a neighborhood).
Existing approaches (including the state-of-the-art 
GPU-based solution~\cite{STIG2016}) implement this query as a
single operator typically making use of a spatial index, which
organizes the minimum bounding rectangle (MBR) of the spatial objects
in a tree structure. The index is used to identify relevant MBRs, and
then, for each point inside the selected MBRs, a test is performed to
check whether the point is indeed inside the query polygon.
Note that this containment test is specific to the input being points. 
If the spatial component of the data is instead represented as the polygon
corresponding to the land plot where the restaurant is located, the
selection query requires a different implementation, since a
polygon-intersect-polygon test must be performed instead.
This shortcoming, coupled with the complexities involved in 
implementing GPU-based solutions, has impeded a wider use
of GPUs in spatial databases.

\myparagraph{Our Approach.}
With the goal of enabling GPUs to be exploited without the need to
build custom solutions, we
revisit the problem of designing a spatial data model and operators.
We adapt common computer graphics
operations for which GPUs are specifically designed and optimized, 
to propose a \emph{new geometric data model that provides a uniform
	representation for different geometric objects}, and an
\emph{algebra consisting of a small set of composable operators
	capable of handling a wide variety of spatial queries}.

While several data models and algebras have been described in the
literature~\cite{Guting88,Samet1995,EGENHOFER91,GARGANO1991,Guting1995},
they were all designed before the advent of modern GPUs and suffer from
at least one of the above shortcomings, as we discuss in
Section~\ref{sec:rel-work}. 
Furthermore, these models are typically \emph{user facing}:
users express the queries of interest by making use of the
data types and the operators provided in the model; the implementation
of the operators is left to the developer.
In contrast, we aim for a \emph{developer-facing} model that
can be incorporated into existing systems unbeknownst to the users,
while at the same time providing significant benefits to the database
engine and query performance.


To give an intuition behind the proposed geometric model, 
consider again the example in Figure~\ref{fig:example}(a) from a 
geometric point of view. The query can be
translated into two operations performed one after the other as shown in
Figure~\ref{fig:example}(b).
Visually (or graphically), the input points and the query
polygon are uniformly represented as \textit{drawings on a canvas}. 
The first operation merges the input points and the query polygon into
a single canvas. The second operation computes the intersection
between the points and the polygon to eliminate points outside the
polygon.
Unlike the traditional execution strategy for spatial aggreagtion
(i.e., join followed by aggregation), the two operations used here are
applicable to any kind of geometry. Therefore, even if the data
(restaurants) were represented as polygons instead of points, the same
set of operations could be applied.

Informally, we represent a spatial object as an 
embedding of its geometry onto a plane, called a \textit{canvas}, 
and define GPU-friendly operators similar to the ones in Figure~\ref{fig:example}(b),
that act on one or more canvases.
As we show in
Section~\ref{sec:expressiveness}, these operators can be re-used
and composed to support a diverse set of spatial
queries. 

Given a small set of basic operators, our model makes it possible for
implementations to focus on the efficiency of these operators, the
gains from which become applicable to a variety of queries. 
While our focus in this paper is on the conceptual representation
and modeling of spatial data and the design of the algebra, 
this work opens new avenues for research in spatial
query optimization, both for theory (\eg developing plan generation
strategies, designing cost models) and systems (\eg designing
indexes leveraging GPUs).

\myparagraph{Contributions.}
To the best of our knowledge, this is the first approach to propose
a query algebra designed with a focus on enabling an efficient GPU realization
of spatial queries.
Our main contributions are as follows:
\begin{itemize}  \denselist
\item We propose a new geometric data model that provides a uniform
    representation for spatial data on GPUs
    (Section~\ref{sec:representation}).
\item We design a spatial algebra consisting of five fundamental
  operators designed based on common computer graphics operations (Section~\ref{sec:operators}). 
We show that the algebra is: expressive and able to represent all standard
spatial queries; 
 and closed, allowing the operators to be composed to
 construct complex queries (Section~\ref{sec:expressiveness}).
\item We present a proof-of-concept implementation of a subset of the 
proposed operators that shows: 
1)~how the proposed model and operators are naturally suited for GPUs; and 
2)~how operators can be re-used in different queries.
Our implementation achieves over two orders of magnitude speedup over
a custom CPU-based implementation, and outperforms 
custom GPU-based approaches (Sections~\ref{sec:implementation} \& \ref{sec:evaluation}).
\item We discuss the compatibility of the proposed algebra 
with the relational model and its utility for query optimization 
(Section~\ref{sec:discussion}).
\end{itemize}

%% file: representation.tex
\begin{figure}[t]
\centering
\includegraphics[width=\linewidth]{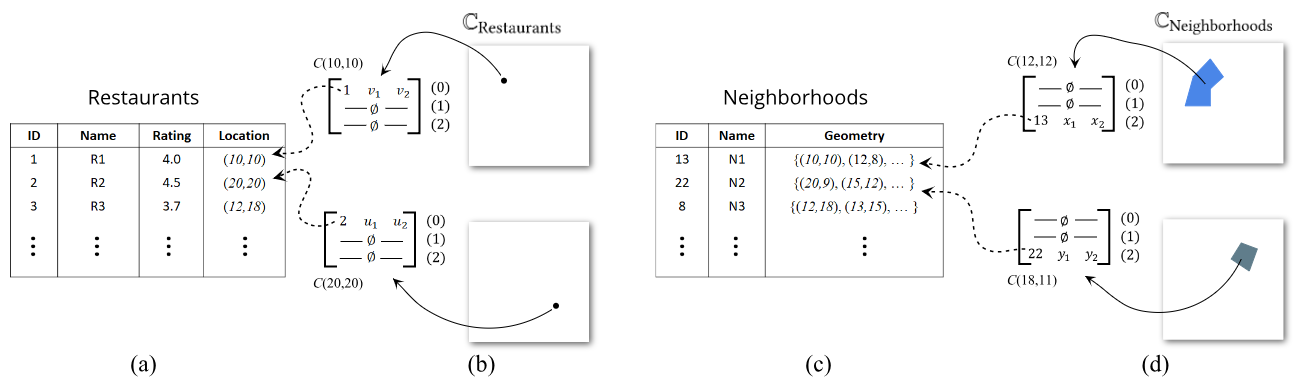}
\vspace{-0.4cm}
\caption{
\textit{Canvas}: A uniform representation of spatial data. 
(a)~Example point data.
(b)~Two canvases corresponding to the first two records of the table. The [0,0] element of the matrix (corresponding to the 0-primitive) stores the unique ID corresponding to the record. 
(c)~Example polygon data.
(d)~Two canvases corresponding to the first two records. Here, all points inside a polygon will map to the same value, with the element [2,0] (corresponding to the 2-primitives) storing the unique ID. 
The white parts of the canvas (not part of the geometry) maps to a null value.
}
\label{fig:canvas-example}
\vspace{-0.2cm}
\end{figure}

\section{Data Representation}
\label{sec:representation}

In this section, we first formalize the notion of a 
spatial data set, and then define the concept of a 
\textit{canvas}, the spatial analogue of a relational tuple.

\subsection{Spatial Data}
\label{sec:data-representation}

As discussed in Section~\ref{sec:intro}, an important limitation of
current spatial operations is that they are tied to specific
representations for geometric data types.
To design flexible operators, we need a uniform representation for the
geometry that is independent of underlying types.
We propose to schematically represent the geometry using a single type
called \textit{geometric object}, which can conceptually represent any
complex geometric structure.

\begin{definition}[Geometric Object]
A geometric object is defined as a collection 
of geometric primitives.
\end{definition}

\begin{definition}[Geometric Primitive]
A $d$-dimensional geometric primitive (\textit{$d$-primitive}) is defined as 
a $d$-manifold (with or without a boundary).
\end{definition}

Informally, a $d$-manifold is geometric space in which the local
neighborhood of every point represents $\R^d$. In the context of
spatial data that is of interest in this work, we focus on
2-dimensional (2D) space and 
$d$-primitives, where $0\leq d\leq 2$.
Intuitively, a 0-primitive is a point while a 
1-primitive is a line (not necessarily a straight line or with a finite length).
2-primitives include any subset of $\R^2$ that is 
neither a line nor a point, such as polygons and half spaces.

A spatial data set can now be defined in terms of geometric objects as follows:
\begin{definition}[Spatial Data]
A spatial data set consists of one or more attributes
of type geometric object.
\end{definition}

Note that the above definition allows geometric objects 
of arbitrarily complex shapes composed using a heterogeneous set that
contains points, lines, as well as polygons. However, geometric objects 
common in real world data sets are primarily only points 
(\eg locations of restaurants, hospitals, bus stops, etc.), 
only lines (\eg road networks), or only polygons (\eg state or city boundaries).

\subsection{Canvas}
\label{sec:canvas}

We define the notion of a canvas to explicitly capture the 
geometric structure of a spatial data set.
As mentioned above, we assume that the dimensions of the
geometric primitives composing a geometric object in a
spatial data set is either 0, 1 or 2.
Let $S$ be a set of $k$-tuples, where $k \geq 1$, such that $\emptyset \in S$.
A \textit{canvas} is formally defined as follows:

\begin{definition}(Canvas)
A canvas is a function $C:\R^2 \rightarrow S^3$ that maps each point in
$\R^2$ to a triple $(s[0],s[1],s[2]) \in (S \times S \times S)$,
where the $i^{th}$ element of the triple, $s[i]$, stores information (as a $k$-tuple)
corresponding to $i$-dimensional geometric primitives.
\end{definition}

\begin{definition}(Empty Canvas)
A canvas $C$ is \textit{empty } iff $C$ maps all points
in $\R^2$ to $(\emptyset,\emptyset,\emptyset)$.
\end{definition}

A canvas is analogous to a tuple in the relational model. 
It is defined such that it captures geometric objects in 
the world coordinate space of the graphics pipeline,
thus making it straightforward to apply computer graphics 
operations on it.
Intuitively, a canvas stores for each point in $\R^2$ information corresponding to
the geometric primitives that pass through that point.
This information is captured by the elements of the set $S$ 
(we discuss $S$ in more detail below).

Given a spatial data set, each record of this data is 
represented using one or more canvases equal to the number
of geometric object attributes of the data.
For ease of exposition, consider a spatial data set having a single
geometric object attribute and a geometric object $o$ corresponding to
one of the records in this data. Let
$o = \{g_1,g_2,g_3,\ldots, g_n\}$, where $g_i$ is a geometric
primitive having dimension $dim(g_i)$, $0 \leq dim(g_i) \leq 2$,
$\forall i$.
A canvas representation of the geometric object $o$ is defined 
as follows. 

\begin{definition}(Canvas representation of a geometric object)
A canvas corresponding to a geometric object is a function 
$C_o:\R^2 \rightarrow S^3$ such that
$\forall d \in [0,2]$
\begin{eqnarray}
C_o(x,y)[d] = 
\begin{cases}
	s_d \neq \emptyset \in S,  & \text{if } \exists i \mid dim(g_i) = d \text{ and } \\
                               & \hspace{0.3in} g_i \text{ intersects } (x,y) \\
	\emptyset & \textrm{otherwise}
\end{cases} \nonumber
\end{eqnarray}
\end{definition}

The set $S$ used in the above definition is called the 
\emph{object information set}, and is defined as follows.

\begin{definition}(Object Information Set $S$)
The object information set $S$ is defined as a set of triples $(v_0,v_1,v_2)$
where $v_0$ stores a unique identifier (or a pointer) for the record corresponding
to the geometric object. $v_1$ and $v_2$ are real numbers
storing meta data related to the canvas.
\end{definition}	
The range of the canvas function $C$ can thus be represented as a
$3 \times 3$ matrix, where each row corresponds to the Object
Information Set for the associated primitive dimension.
We abuse notation to represent the triple $(\emptyset, \emptyset, \emptyset)$ simply as $\emptyset$.
\begin{figure}[t]
	\centering
	\includegraphics[width=0.6\linewidth]{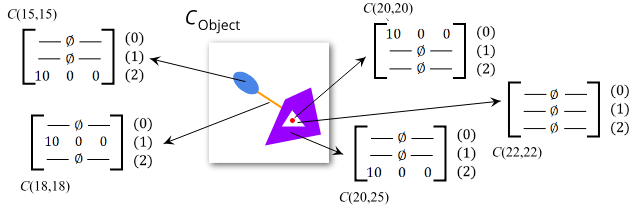}
	\vspace{-0.4cm}
	\caption{
		A canvas representing a complex object.
		Since all the primitives (colored differently) are part of the same object, they have
		the same ID.
	}
	\label{fig:canvas-complex}
	\vspace{-0.4cm}
\end{figure}

\begin{figure}[t]
	\centering
	\includegraphics[width=\linewidth]{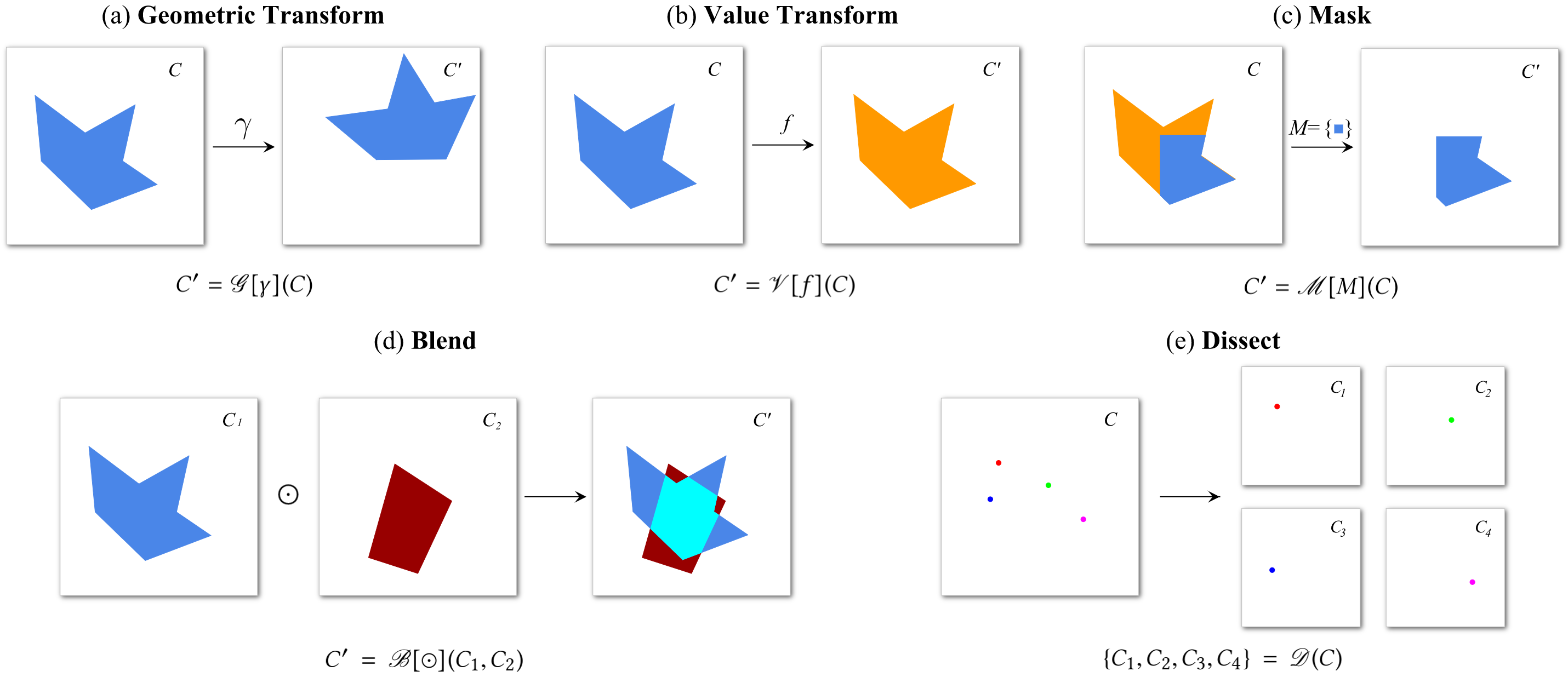}
	\vspace{-0.2in}
	\caption{The 5 fundamental operators. For illustrative purposes, we use colors to denote the information
		stored in each point of the canvas; the white color corresponds to a null value.}
	\label{fig:operators}
	\vspace{-0.3cm}
\end{figure}

\myparagraph{Example 1:}
Consider the two example \datasets in Figure~\ref{fig:canvas-example}.
The first data set corresponds to the set of restaurants in a city
(represented as points)~(a),
while the second data set corresponds to the neighborhood boundaries
of this city (represented as polygons)~(c).
Figure~\ref{fig:canvas-example} also illustrates the canvas representations
corresponding to two records from each of these two \datasets.
Note that in this example, we use only the identifier element of the 
object information set. The values of the other elements are initialized
depending on the query scenario (Section~\ref{sec:expressiveness}).

\myparagraph{Example 2:}
The complex geometric object shown in Figure~\ref{fig:canvas-complex} 
consists of two polygons (an ellipse and a polygon with a hole) 
connected by a line, with the hole also containing a point.
This is represented in the canvas by mapping the regions
corresponding to the different primitives using the
appropriate rows in the matrix (for point, line, polygon).

%% file: operators.tex
\section{Operators}
\label{sec:operators}

Below, we define the operators we designed for the canvas
representation of a spatial data set. Concrete examples of how these
operators are used are given in Section~\ref{sec:expressiveness}. We categorize the set of
operators into three classes: fundamental, derived, and utility
operators.
We use the following notation to represent operators that take 
as input zero or more canvases:
\begin{center}
$Op[P_1,P_2,\ldots](C_1,C_2,\ldots,C_n)$
\end{center}
where $Op$ is the operator name, $P_i, \forall i$, the parameters of the operator,
and $C_j, \forall j$, the canvases input to the operator. 
The output of all the operators is always
one or more canvases. Thus, \textit{the proposed algebra is closed by
  design}.

\subsection{Fundamental Operators}
\label{sec:op-fundamental}

Fundamental operators form the core of the 
proposed algebra. Their design is based on common computer graphics
operations that are already supported in GPUs.
Figure~\ref{fig:operators} illustrates the five fundamental
operators.

\myparagraph{Geometric Transform $C' = \opG[\gamma](C)$:}
This operator takes as input a single canvas $C$ and outputs a canvas $C'$
in which all the geometric objects of $C$ are transformed to a new position
in $C'$ defined by the parameter function $\gamma$.
Here, the parameter function $\gamma$ can be defined in two ways:
\begin{enumerate}\denselist
\item $\gamma:\R^2 \rightarrow \R^2$
\item $\gamma:S^3 \rightarrow \R^2$
\end{enumerate}

\noindent In the first case, the new position $(x',y')$ of a geometry is dependent on its 
current position $(x,y)$:
\[ C'(\gamma(x,y)) = C(x,y)\]
Examples of such functions include operations such as rotation, translation, etc.
The example in Figure~\ref{fig:operators}(a) rotates and translates (moves) the
polygon object to a different position.

A scenario where this operator is useful is when 
spatial \datasets 
in a database use different coordinate systems. 
Thus, when performing binary or n-ary operations on canvases from these 
\datasets, the geometry has to be converted into a common coordinate system first.
The parameter function $\gamma$ can be defined appropriately for this purpose.

In the second case, the new position $(x',y')$ of the geometry is dependent on 
the information stored at the current position $C(x,y)$:
\[ C'(\gamma(C(x,y))) = C(x,y)\]
Such a transformation is useful, for example, when one is interested in accumulating
values (\eg for aggregation queries) corresponding to a geometric object---in this case, the function $\gamma$ can be
defined to move all points having the same object identifier to a unique location.

\myparagraph{Value Transform $C' = \opV[f](C)$:}
This unary operator outputs a canvas $C'$
in which the information corresponding to the geometries is modified 
based on the parameter function $f$. That is,
\[ C'(x,y) = f(x,y,C(x,y))\]
where, $f:\R^2 \times S^3 \rightarrow S^3$ is a function that changes the
object information based on its location and/or value.
Figure~\ref{fig:operators}(b) illustrates an example of this 
operation where the color of the polygon in the canvas is changed
from blue to orange. 

\myparagraph{Mask $C' = \opM[M](C)$:}
The mask operator is used to filter regions of canvas so that only regions satisfying 
the condition specified by $M \subset S^3$ are retained. Formally, the application of this operator
results in the canvas $C'$ such that
\[
C'(x,y) = 
\begin{cases}
	C(x,y), & \text{if~~} C(x,y) \in M \\
	\emptyset & \text{otherwise}
\end{cases}
\]
For example, this can be used to
accomplish select intersection operations shown 
in Figure~\ref{fig:example}(b) and Figure~\ref{fig:operators}(c). 

\myparagraph{Blend $C' = \opB[\odot](C_1,C_2)$:}
Blend is a binary operator used to merge two canvases into
one. The blend function $\odot$:$S^3 \times S^3 \rightarrow S^3$ 
defines how the merge is performed:
\[C'(x,y) = C_1(x,y) \odot C_2(x,y)\]
The merge operation used in Figure~\ref{fig:example}(b) is an 
instance of the blend function.  Another example is shown in Figure~\ref{fig:operators}(d).

\myparagraph{Dissect $\{C_1,C_2,\ldots,C_n\} = \opD(C)$:}
The dissect operation splits a given canvas into 
multiple non-empty canvases, each corresponding to a point 
$(x,y) \in \R^2$ having $C(x,y) \neq \emptyset$.
That is, a new canvas $C_i$ is generated corresponding 
to a non-null point $(x,y)$ such that
\[
C_i(x',y') = 
\begin{cases}
	C(x,y), & \text{if~~} (x',y') = (x,y) \\
	\emptyset & \text{otherwise}
\end{cases}
\]
For example, in Figure~\ref{fig:operators}(e), a canvas encoding 4 points is 
split into 4 canvases each corresponding to one of those points.
As we show later, one of the uses of this operator is for queries involving
aggregations over geometries with 1- and 2-primitives (such as
polygons).

\subsection{Derived Operators}
\label{sec:op-derived}

It is common for certain combinations of fundamental operators
to be repeatedly used for various queries. We represent these
combinations as derived operators and describe of couple of 
such useful operators below.

\myparagraph{Multiway Blend $C' = \opB^*[\odot](C_1,C_2,\ldots,C_n)$:}
This n-ary operator takes as input $n$ canvases and 
generates a single canvas after blending all these $n$ canvases in the given order.
\[
C' = \opB[\odot](C_1, \opB[\odot](C_2, \opB[\odot](C_3, \ldots))) 
\]
Note that if the blend function $\odot$ is associative, then it allows
relaxing the grouping of the different blend operations, thus providing more
flexibility while optimizing queries.

\myparagraph{Map $\{C_1,C_2,\ldots,C_n\} = \opD^*[\gamma](C)$:}
Map is a composition of a dissect
followed by a geometric transform.
\itemspace
\[
\{C_1,C_2,\ldots,C_n\} = \opG[\gamma](\opD(C))
\]
This operator is useful to align all the canvases
resulting from the dissect. In such a case, $\gamma$
is typically defined as a constant function:
\itemspace \itemspace
\[
\gamma(x,y) = (x_c,y_c)
\]
where $x_c$ and $y_c$ are constants.

Without loss of generality, we assume the above notation 
of providing multiple canvases as input 
to a unary operator (in this case the geometric transform
that takes as input canvases output from a dissect operation) 
as equivalent to applying the operator 
individually to each of the input canvases. 

\subsection{Utility Operators}
\label{sec:op-utility}

Utility operators are primarily used to generate canvases
based on a set of parameters. These are particularly useful 
for classes of spatial queries involving parametric constraints.
In particular, we consider the following three types of
utility operators:

\myparagraph{Circle $C = Circ[(x,y),r]()$:}
generates a canvas corresponding to a circle 
with center is $(x,y)$ and radius $r$.

\myparagraph{Rectangle $C = Rect[l_1,l_2]()$:}
generates a canvas corresponding to a rectangle
having diagonal end points $l_1$ and $l_2$.

\myparagraph{Half Space $C = HS[a,b,c]()$:}
generates a canvas representing the half space
defined by the equation: $ax + by + c < 0$.

%% file: queries.tex
\section{Expressiveness}
\label{sec:expressiveness}

To demonstrate the expressiveness of the proposed model, in what follows
we describe how common spatial queries can be represented as algebraic 
expressions. 
We build upon the classification of spatial queries used by 
Eldawy et al.~\cite{Eldawy2016} and
categorize spatial queries into the following classes:
\textit{selection}, \textit{join}, \textit{aggregate},
\textit{nearest neighbor}, and \emph{geometric queries}.
Note that this is a super set of the query types
evaluated in a state-of-the-art experimental survey
by Pandey~et~al.~\cite{pandey@vldb2018}.

For ease of exposition, we consider only point and polygonal data sets.
It is straightforward to express similar queries for other types of 
spatial data sets with lines, or more complex geometries (combination of
points, lines and polygons).
Without loss of generality, we assume that the different operators
prune empty canvases from their output, and an empty canvas is generated
when an input canvas does not satisfy a given constraint. 
This is similar to the relational model, where tuples that do not
match the query constraints are excluded from the output table.

\subsection{Selection Queries}
\label{sec:select-queries}

We can classify spatial selection queries into 
three types: polygonal selection, range selection, and
distance-based selection.\footnote{While nearest-neighbor-based
selection could also be in this category, we
place it in a separate class (Section~\ref{sec:nn-queries}).}
We first consider selection queries that have a polygonal 
constraints, and then extend the algebraic expressions 
for other types of selection queries.

\begin{figure}[t]
	\centering
	\includegraphics[width=\linewidth]{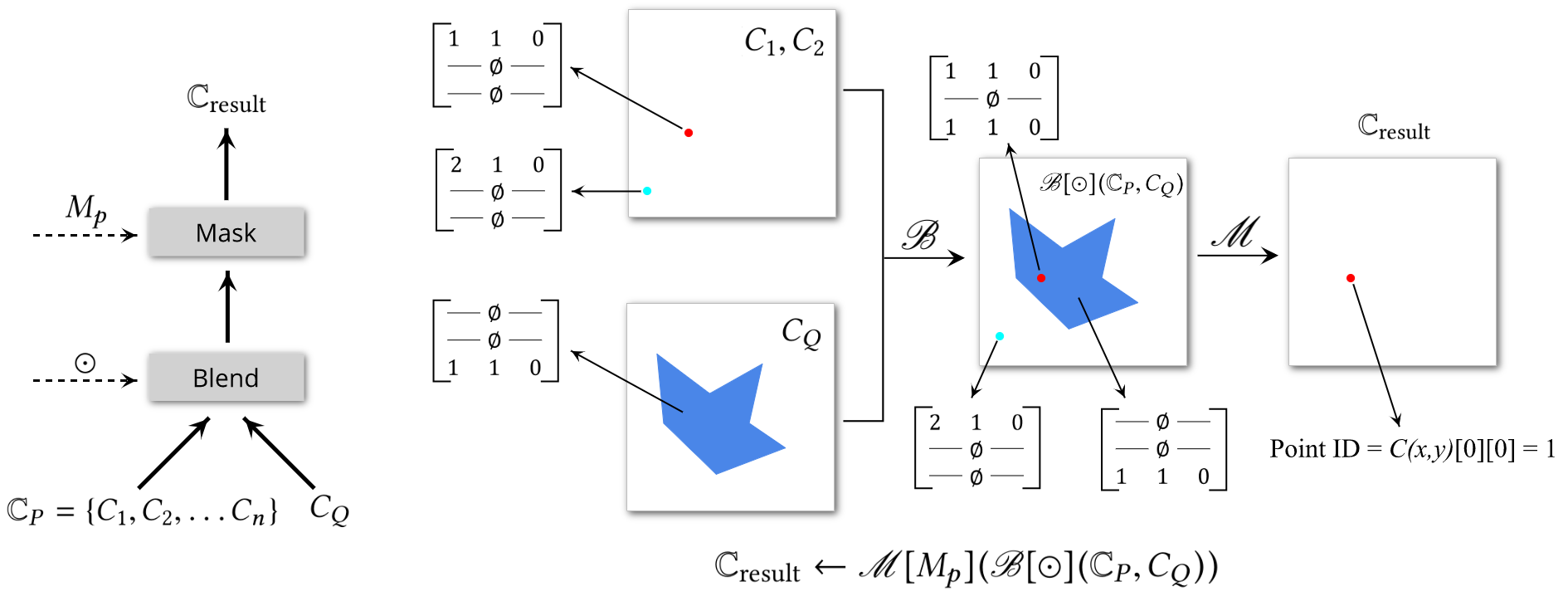}
	\vspace{-0.15in}
	\caption{A schematic representation as a plan diagram of the algebraic expression 
		used to select points based on a polygonal constraint (left). The different steps
		of this plan is illustrated 
		using an example input with 2 points (colored red and cyan)(right).
		For simplicity, both points are shown in a single canvas.}
	\label{fig:exp-select-points}
	\vspace{-0.4cm}
\end{figure}

\myparagraph{Polygonal Selection of Points.}
Let $D_P$ be a data set consisting of a set of points. 
Let $\left\lbrace (x_1,y_1), (x_2,y_2), \ldots, (x_n,y_n) \right\rbrace$
be the coordinates corresponding to the location of these points.
Let $Q$ be any arbitrary-shaped polygon.
Consider the following spatial query expressed in an SQL-like syntax:

\noindent 
\begin{boxedminipage}[b]{\linewidth}{}
\centering
\small
\texttt{SELECT * FROM $D_P$ WHERE Location INSIDE $Q$}
\end{boxedminipage}

\sloppy
Note that this is the same query used for the example in Figure~\ref{fig:example}(a).
Using the proposed data representation, let
$\C_P = \{C_1, C_2, \ldots C_n\}$ be the set of canvases
corresponding to each point (record) in $D_P$.
Let the canvas $C_i$ corresponding to the $i^{\text{th}}$ record be defined as follows:
\itemspace
\begin{eqnarray}
C_i(x,y)[0] &=&  
\begin{cases}
(id,1,0) & \text{if  } (x,y) = (x_i,y_i) \nonumber \\ 
\emptyset & \text{otherwise}
\end{cases} \nonumber \\ 
C_i(x,y)[1] &=& \emptyset \nonumber \\ 
C_i(x,y)[2] &=& \emptyset \nonumber
\end{eqnarray}
Here, $id$ corresponds to the unique identifier mapping the canvas to
the corresponding record in $D_p$.
We use the second element of the tuple $C_i(x,y)[0]$ to keep count of the points
incident on the location $(x,y)$, which in this case is $1$. The third element
is ignored for this query.
Let the canvas $C_Q$ corresponding to the query polygon $Q$ be defined as follows:
\begin{eqnarray}
C_Q(x,y)[0] &=& \emptyset \nonumber \\ 
C_Q(x,y)[1] &=& \emptyset \nonumber \\ 
C_Q(x,y)[2] &=&  
\begin{cases}
(1,1,0) & \text{if  } (x,y) \text{ falls inside } Q \\ \nonumber
\emptyset & \text{otherwise}
\end{cases} \nonumber
\end{eqnarray}
Similar to the case of points above, the elements $C_Q(x,y)[2][0]$
and $C_Q(x,y)[2][1]$ stores the $id$ of the query polygon 
(set to $1$ since there is only one polygon) 
and count of 2-primitives incident on $(x,y)$ respectively.
Using the canvases defined above, the select query can be algebraically expressed as follows:
\[
 \C_{\text{result}} \leftarrow \opM[M_p](\opB[\odot](\C_P,C_Q))
\]
where, $\forall s_1, s_2 \in S^3$
\[
s_1 \odot s_2 = 
\begin{bmatrix}
s_1[0][0] & s_1[0][1] & s_1[0][2] \\
\text{---} & \emptyset & \text{---} \\
s_2[2][0] & s_2[2][1] & s_2[2][2]
\end{bmatrix}
\]
and
\[
M_p = \left\lbrace s \in S^3 \mid s[0] \neq \emptyset \text{ and } s[2][0] = 1 \right\rbrace 
\]

\begin{figure}[t]
	\centering
	\includegraphics[width=\linewidth]{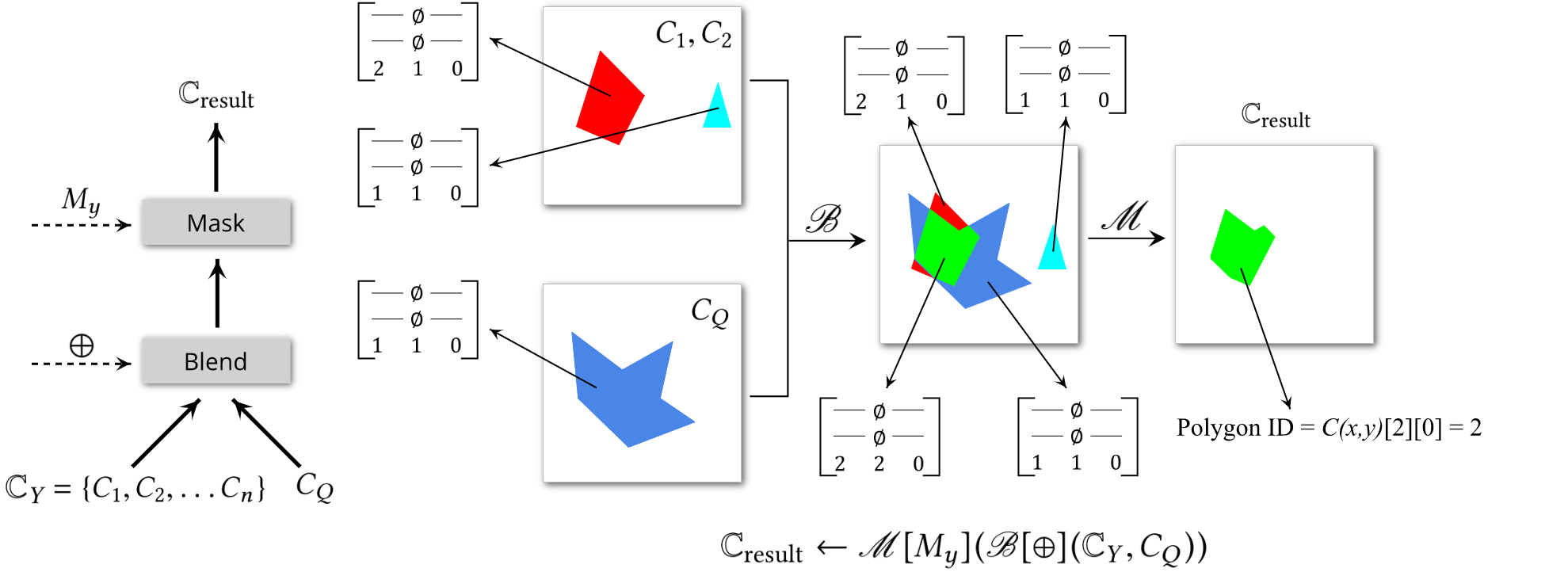}
	\vspace{-0.2in}
	\caption{Plan diagram corresponding to the algebraic expression 
		used to select polygons based on a polygonal constraint (left). This
		plan is illustrated 
		using 2 input polygons (colored red and cyan) (right), which
		are shown in a single canvas.}
	\label{fig:exp-select-polys}
	\vspace{-0.5cm}
\end{figure}

Similar to the example in Figure~\ref{fig:example}(b),
the above expression first merges the input data with 
the query polygon using the blend operator $\opB$,
and then uses the mask operator $\opM$ to select only the
intersection (a location is part of the intersection
if both, a 1-primitive and 2-primitive are incident on it).
Figure~\ref{fig:exp-select-points} visualizes the above expression as a 
plan diagram, and illustrates the different steps for two examples when a point
is inside the query polygon (and hence part of the result), 
and when a point is outside respectively.
 
\myparagraph{Polygonal Selection of Polygons.}
Let $D_Y$ be a data set consisting of a set of polygons. 
Let $\left\lbrace Y_1, Y_2, \ldots, Y_n \right\rbrace$ be the
set of polygons associated with each record of the data set.
As before, the polygons can take any shape.
Let $Q$ be another arbitrary-shaped polygon.

Let the set of canvases $\C_Y$ corresponding to polygons in $D_Y$ be defined as follows:
\begin{eqnarray}
C_i(x,y)[0] &=& \emptyset \nonumber \\ 
C_i(x,y)[1] &=& \emptyset \nonumber \\
C_i(x,y)[2] &=&  
\begin{cases}
(id,1,0) & \text{if  } (x,y) \text{ falls inside } Y_i \nonumber \\ 
\emptyset & \text{otherwise}
\end{cases} \nonumber 
\end{eqnarray}
Let the canvas corresponding to query polygon $Q$ be defined as before.
Now, consider the following selection query, similar to the one above, 
but over $D_Y$:
\noindent 
\begin{boxedminipage}[b]{\linewidth}{}
\centering
\small
\texttt{SELECT * FROM $D_Y$ WHERE Geometry INTERSECTS $Q$}
\end{boxedminipage}

\noindent This query can be algebraically expressed as follows:
\[
 \C_{\text{result}} \leftarrow \opM[M_y](\opB[\oplus](\C_Y,C_Q))
\]
where, $\forall s_1, s_2 \in S^3$
\[
s_1 \oplus s_2 = 
\begin{bmatrix}
\text{---} & \emptyset & \text{---} \\
\text{---} & \emptyset & \text{---} \\
s_1[2][0] & s_1[2][1] + s_2[2][1] & s_1[2][2]
\end{bmatrix}
\]
and
\[
M_y = \left\lbrace s \in S^3 \mid s[2][1] = 2 \right\rbrace 
\]

Note that unlike in the previous case of selecting points, 
since both the data as well as the query consist of
polygons, both the data canvas and the query canvas store
information only for 2-primitives. Hence, the second element of
the information tuple is made use of in this case to
compute the intersection (\ie locations having two 2-primitives
incident on them).
Figure~\ref{fig:exp-select-polys} shows the algebraic expression
using a plan diagram, and illustrates two examples
denoting selection and non-selection scenarios respectively.

\myparagraph{Selection Using Other Spatial Constraints.}
In addition to polygonal constraints, selection queries over
spatial data may also involve other types of spatial constraints.
Commonly used are range constraints and distance-based selection.
It is easy to extend the algebraic expressions used for
polygonal constraints to these scenarios as follows.

\myparagraphem{1. Rectangular Range Constraints:} 
This class of queries requires the selection of spatial objects that 
intersect a 2D range. To execute such queries,
the query polygon is simply replaced by a rectangle, the canvas
for which can be created using the utility operator:
$C_Q \leftarrow Rect[l_1,l_2]()$, 
where $l_1,l_2$ denotes the diagonal endpoints of the rectangle range.

\myparagraphem{2. One-Sided Range Constraints:} 
In this scenario, the queries require 
selecting geometries that intersect a given half-space $ax+by+c < 0$
(note that this is a more generic formulation of queries
involving constraints such as $x < c$ or $y < c$).
Again, the utility operator can be used to generate the
required query canvas as a replacement for the query polygon: \\
$C_Q \leftarrow HS[a,b,c]()$.

\myparagraphem{3. Distance-based Selection:}
In this case, the queries require the selection of geometries that lie within a given 
distance $d$ of a query point $(x_q,y_q)$. This essentially translates to 
using a circle with radius $d$ centered at $(x_q,y_q)$ as the query polygon,
the canvas for which can also be created using the utility operator:
$C_Q \leftarrow Circ[(x_q,y_q),d]()$

\smallskip
Given the possibility to adapt these three types of spatial
constraints to a polygon, we will focus only on polygonal constraints 
for the remainder of this section.

\subsection{Join Queries}
\label{sec:join-queries}

Spatial join queries can be broadly classified into three types:
(Type~I)~points $\bowtie$ polygons join;
(Type~II)~polygons $\bowtie$ polygons join; and
(Type~III)~points $\bowtie$ points join.
Type~III is commonly known as a distance join. 
As in the previous section, one set of points (say the RHS) of the distance
join can be converted into a collection of circles to transform 
this to a points $\bowtie$ polygons join query.
We therefore focus on the first two types of join queries.

Let $D_P$ and $D_Y$ be a point data set and a polygon data set respectively.
A Type~I join query between these two data sets is typically specified as follows:

\noindent 
\begin{boxedminipage}[b]{\linewidth}{}
\centering
\small
\texttt{SELECT * FROM $D_P, D_Y$ WHERE $D_P.$Location INSIDE $D_Y$.Geometry}
\end{boxedminipage}

\newpage
\noindent Similarly, let $D_{Y1}$ and $D_{Y2}$ be two polygon data sets. A
Type~II join query between these two data sets can be specified as follows:

\noindent 
\begin{boxedminipage}[b]{\linewidth}{}
\centering
\small
\texttt{SELECT * FROM $D_{Y1}, D_{Y2}$ WHERE $D_{Y1}$.Geometry INTERSECTS $D_{Y2}$.Geometry}
\end{boxedminipage}

\noindent The above join queries
are equivalent to performing selection queries, one for each record (canvas) 
from $D_Y$ and $D_{Y2}$ respectively.
Thus, conceptually, the algebraic expression for joins
is the same as the corresponding selection queries, with the exception
that a single query polygon is instead replaced with a collection
of polygons.
A Type~I join query can then be realized using the following expression
\[
 \C_{\text{result}} \leftarrow \opM[M_p](\opB[\odot](\C_P,\C_Y))
\]
while a Type~II join query can be realized using
\[
 \C_{\text{result}} \leftarrow \opM[M_y](\opB[\oplus](\C_{Y1},\C_{Y2}))
\]
Here, $\C_P$, $\C_Y$, $\C_{Y1}$, and $C_{Y2}$ are collections of canvases
corresponding to the data sets $D_P$, $D_Y$, $D_{Y1}$ and $D_{Y2}$ respectively.
The different parameters of the operators in the above expressions remain the same
as what was used for their selection counterparts.

Similar to the join operator in the relational model, spatial joins using
our model can be implemented in several ways. A
straightforward approach is using nested loops for the blend
operation, which can be made more efficient if  spatial indexes 
are available.

\subsection{Aggregate Queries}
\label{sec:aggregate-queries}

The third class of queries common for spatial data are spatial aggregation queries.
We consider two types of such queries: aggregating the results 
from a selection, and the aggregation required for a group-by over a join.

\myparagraph{Aggregation over a Select.}
Consider first a simple count of the results from a selection query:

\noindent 
\begin{boxedminipage}[b]{\linewidth}{}
\centering
\small
\texttt{SELECT COUNT(*) FROM $D_P$ WHERE Location INSIDE $Q$}
\end{boxedminipage}

\noindent This query can be realized using the expression:
\[
 C_{\text{count}} \leftarrow \opB^*[+](\opG[\gamma_c](\C_{\text{result}}))
\]
where $\gamma_c:S^3 \rightarrow \R^2$ is defined such that
\[
\forall s \in S^3, \gamma_c(s) = (s[2][0],0), 
\]
$+:S^3 \times S^3 \rightarrow S^3$ is defined as
\[
s_1 + s_2 = 
\begin{bmatrix}
0 & s_1[0][1] + s_2[0][1] & 0 \\
\text{---} & \emptyset & \text{---} \\
s_2[2][0] & s_2[2][1] & s_2[2][2] \\
\end{bmatrix}
\]
and
\[
 \C_{\text{result}} \leftarrow \opM[M_p](\opB[\odot](\C_P,C_Q))
\]
is the set of canvases resulting from the select operation 
(same as in Section~\ref{sec:select-queries} above).

\begin{figure}[t]
	\centering
	\includegraphics[width=0.95\linewidth]{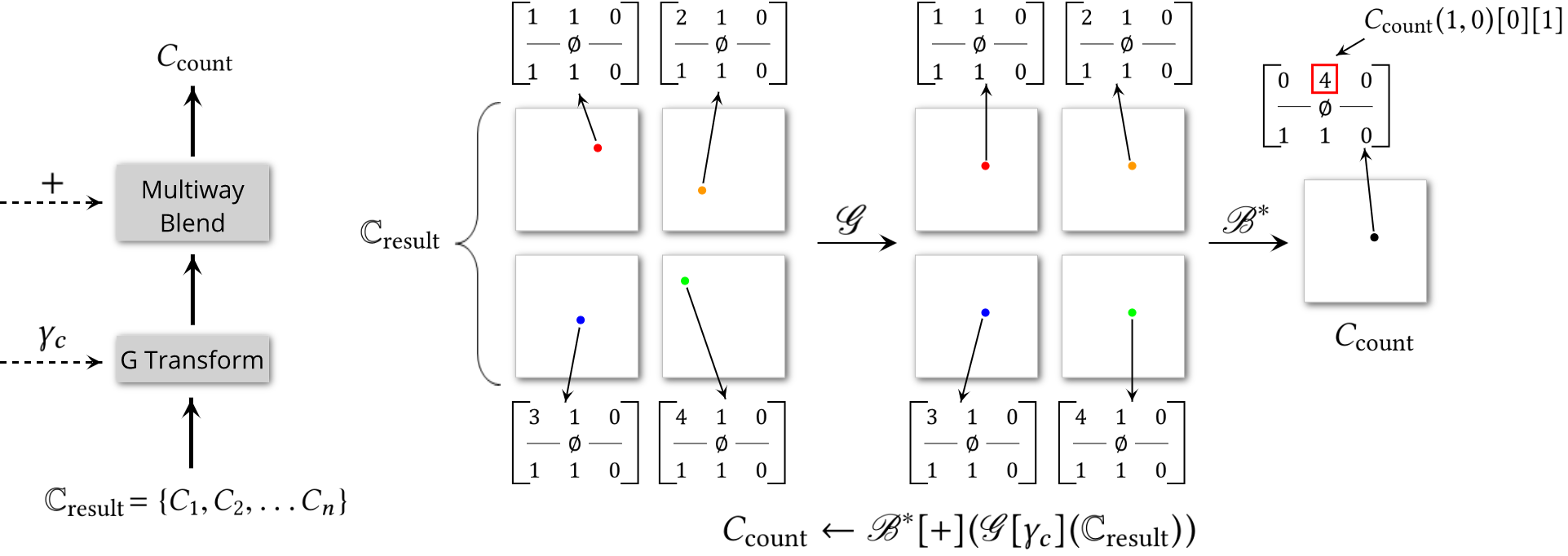}
	\vspace{-0.2in}
	\caption{Plan diagram for aggregating the results from a select query (left). 
		The example (right) uses the results from a select sub-query 
		that returns 4 points, and illustrates the workflow
		that counts the results.}
	\label{fig:exp-select-count}
	\vspace{-0.3cm}
\end{figure}

Basically, each canvas (corresponding to a point) satisfying the
selection constraint is transformed to a constant location $(1,0)$
(recall that the $id$ of the query polygon $Q$ is 1),
and the resulting canvases are merged together to compute
the required summation (see Figure~\ref{fig:exp-select-count}). 
The value of $C_{\text{count}}(1,0)[0][1]$
stores the resulting count. Note that the second element of the
tuple corresponding to the 0-primitives is used for this operation,
while this was not necessary when performing only a select.

Instead of count, if the query requires computing other 
distributive (\eg sum, minimum, maximum)
or algebraic (\eg average) aggregations over a given attribute, then the third element of
the tuple corresponding to the 0-primitives can be used to store the 
value corresponding to this attribute, and the $+$ function can
be modified appropriately.
For example, let $A$ be a real-valued attribute of the data set $D_P$.
Consider the following query:

\noindent 
\begin{boxedminipage}[b]{\linewidth}{}
\centering
\small
\texttt{SELECT SUM($A$) FROM $D_P$ WHERE Location INSIDE $Q$}
\end{boxedminipage}

This query can be realized using the same expression as above by defining
$C_i \in \C_P$ and $+$, respectively, as follows.
\begin{eqnarray}
C_i(x,y)[0] &=&  
\begin{cases}
(id,1,A[i]) & \text{if  } (x,y) = (x_i,y_i) \nonumber \\ 
\emptyset & \text{otherwise}
\end{cases} \nonumber \\ 
C_i(x,y)[1] &=& \emptyset \nonumber \\ 
C_i(x,y)[2] &=& \emptyset \nonumber
\end{eqnarray}
\[
s_1 + s_2 = 
\begin{bmatrix}
0 & s_1[0][1] + s_2[0][1] & s_1[0][2] + s_2[0][2] \\
\text{---} & \emptyset & \text{---} \\
s_2[2][0] & s_2[2][1] & s_2[2][2] \\
\end{bmatrix}
\]
In this scenario, the value of $\C_{\text{result}}(1,0)[0][2]$
maintains the required sum.

\myparagraph{Aggregation over a Join.}
The second type of aggregation queries consist of a group-by operation 
over a spatial join. In particular, consider the following query:

\noindent 
\begin{boxedminipage}[b]{\linewidth}{}
\centering
\small
\texttt{SELECT COUNT(*) FROM $D_P, D_Y$
WHERE $D_P.$Location INSIDE $D_Y.$Geometry\\
~~~~~~GROUP BY $D_Y.$ID}
\end{boxedminipage}

\noindent The algebraic expression used for aggregations over select works for this query as well:
\[
 C_{\text{count}} \leftarrow \opB^*[+](\opG[\gamma_c](\C_{\text{result}}))
\]
where 
\[
 \C_{\text{result}} \leftarrow \opM[M_p](\opB[\odot](\C_P,\C_Y))
\]

When using the expression for a join, 
each of the polygons have a unique $id$. Hence, the join result corresponding to
a point-polygon pair that satisfies the containment constraint
will be moved to the location $(id,0)$ corresponding to that polygon.
Thus, the final multiway blend operation will individually count points
within each of the polygons in $D_Y$. The value $C_{\text{count}}(id,0)[0][1]$
stores the value corresponding to polygon with ID $id$.

\subsection{Nearest-Neighbor Queries}
\label{sec:nn-queries}

We consider the following nearest-neighbor-based query template that
finds the $k$ points closest to a given query point $X(x_p,y_p)$
($k$NN query).

\noindent 
\begin{boxedminipage}[b]{\linewidth}{}
\centering
\small
\texttt{SELECT * FROM $D_P$ WHERE Location $\in$ KNN($X$, $k$)}
\end{boxedminipage}
Without loss of generality, we assume that the distances of points
in $D_P$ to query point $X$ are totally ordered, \ie no two distances
are the same. In the presence of a clash, the points can be perturbed 
by an infinitesimally small distance $\epsilon$ to ensure the total order
condition is satisfied.

One way to answer this query is to first find the distance $r$ such that
there are exactly $k$ points within the circle centered at $X$ with
radius $r$. Then, the distance-based selection can be used
to obtain the query result. This workflow can be accomplished
using the proposed algebra as follows.
Let $\C_X$ be a set of circles centered at $X$ 
have increasing radii.\footnote{Conceptually there is an
infinite number of circles, but in practice, a finite number 
of circles can be created with small increments in 
radii up to a maximum radius.} This can be accomplished by
using the $Circ()$ utility operator.
Let the $id$ of each circle $c$ be the radius of $c$.
Then, the required radius $r$ to identify the $k$ nearest neighbors 
can be obtained using the following expression:
\[
 \C_{r} \leftarrow \opD^*[\gamma_0](\opM[M_r](C_{\text{count}})) \textrm{~where}
\]
\[
M_r = \left\lbrace s \in S^3 \mid s[0][1] = k \right\rbrace
\]
$\gamma_0:S^3 \rightarrow \R^2$ is defined as
\[
\forall s \in S^3, \gamma_0(s) = (0,0) \textrm{, and}
\]
\[
C_{\text{count}} \leftarrow \opB^*[+](\opG[\gamma_c](\opM[M_p](\opB[\odot](\C_P,\C_X))))
\]
is the same join-group-by aggregation used above.
Essentially, the \textit{mask} operation is applied to the result from the 
aggregation to remove all circles containing fewer 
or more than $k$ points, followed by a \textit{map} to
obtain individual canvases for each valid radius.
Therefore, $C(0,0)[2][0], \forall C \in \C_r$ has the $id$s of canvases 
corresponding to the circles having exactly $k$
points in them. Since the $id$s correspond to the radius of the respective circles,
this can then be used to perform a distance-based selection
to complete the $k$NN query.

\subsection{Computational Geometry Queries}
\label{sec:geom-queries}

The final class of queries we consider is the 
set of computational geometry queries. These include
queries such as computing the Voronoi diagram, spatial skyline, and
convex hull~\cite{Eldawy2016}.
While it might not be straightforward to realize all of these
queries algebraically, the provided operators can be used 
as part of a stored procedure to execute some of them.

We illustrate this through the following example: consider the query to compute the Voronoi diagram
for a given set of points $\left\lbrace (x_1,y_1), (x_2,y_2), \ldots, (x_n,y_n) \right\rbrace$.
This can be accomplished using the following pseudo-code:
\newlength{\oldtextfloatsep}\setlength{\oldtextfloatsep}{\textfloatsep}
\setlength{\textfloatsep}{0pt}
\vspace{-0.2cm}
\begin{procedure}[h]
\small
\caption{ComputeVoronoi()}
\label{code:voronoi}
\begin{algorithmic}[1]
\REQUIRE Points $\left\lbrace (x_1,y_1), (x_2,y_2), \ldots, (x_n,y_n) \right\rbrace$
\STATE $C_\text{voronoi} \leftarrow \emptyset$ 
\FOR{each $i \in [1,n]$}
	\STATE $C_\text{voronoi} \leftarrow \opV[f_{(x_i,y_i)}](C_\text{voronoi})$
\ENDFOR
\RETURN $C_\text{voronoi}$
\end{algorithmic}
\end{procedure}

\vspace{-0.2cm}
\noindent Here, $f_{(x_p,y_p)}:\R^2 \times S^3 \rightarrow S^3$ is defined as follows:
\begin{eqnarray}
f_{(x_p,y_p)}(x,y,s)[0] &=& \emptyset \nonumber \\ 
f_{(x_p,y_p)}(x,y,s)[1] &=& \emptyset \nonumber \\ 
f_{(x_p,y_p)}(x,y,s)[2] &=&
\begin{cases}
(i,d_2,0) & \text{if } s = \emptyset \nonumber \\ 
(s[2][0], s[2][1], 0) & s[2][1] < d_2 \nonumber \\
(i,d_2,0) & \text{otherwise} \nonumber
\end{cases} \nonumber 
\end{eqnarray}
where $d_2$ is the Euclidean 
distance between the point $(x,y)$ and the parameter point $(x_p,y_p)$.
This procedure incrementally builds the Voronoi diagram
by adding one input point at a time---during iteration $i$, the
regions of existing polygons closest to point $i$ are merged to form a
new Voronoi region corresponding to this point. 

\setlength{\textfloatsep}{\oldtextfloatsep}
It might not be possible to express all
computational geometry queries as stored procedures
using the previously defined operators. In such cases
new operators can be added for such queries.

\subsection{Complex Queries}
\label{sec:complex-queries}

So far we focused on standard queries and showed
how they could be translated into algebraic expressions
using the proposed algebra.
As mentioned in Section~\ref{sec:intro}, an algebra is useful
only if the operators can be easily composed to also support 
more complex queries. 
In this section, we demonstrate this property using 
a spatial query involving constraints on
two spatial attributes~\cite{taxivis}: we consider selection queries over
origin-destination data sets (e.g., taxi trips, migration data), where the selection 
is based on polygonal constraints on both origin as
well as destination locations:

\noindent 
\begin{boxedminipage}[b]{\linewidth}{}
\centering
\small
\texttt{SELECT * FROM $D_P$
WHERE Origin INSIDE $Q_1$ and Destination INSIDE $Q_2$}
\end{boxedminipage}
Here, $D_P$ is the input point data set having two location attributes
\textit{Origin} and \textit{Destination}, and 
$Q_1$ and $Q_2$ are polygonal constraints over the two
location attributes respectively. 
This query could be used, for example, to retrieve all the taxi trips
between two specific neighborhoods.

\begin{figure}[t]
	\includegraphics[width=\linewidth]{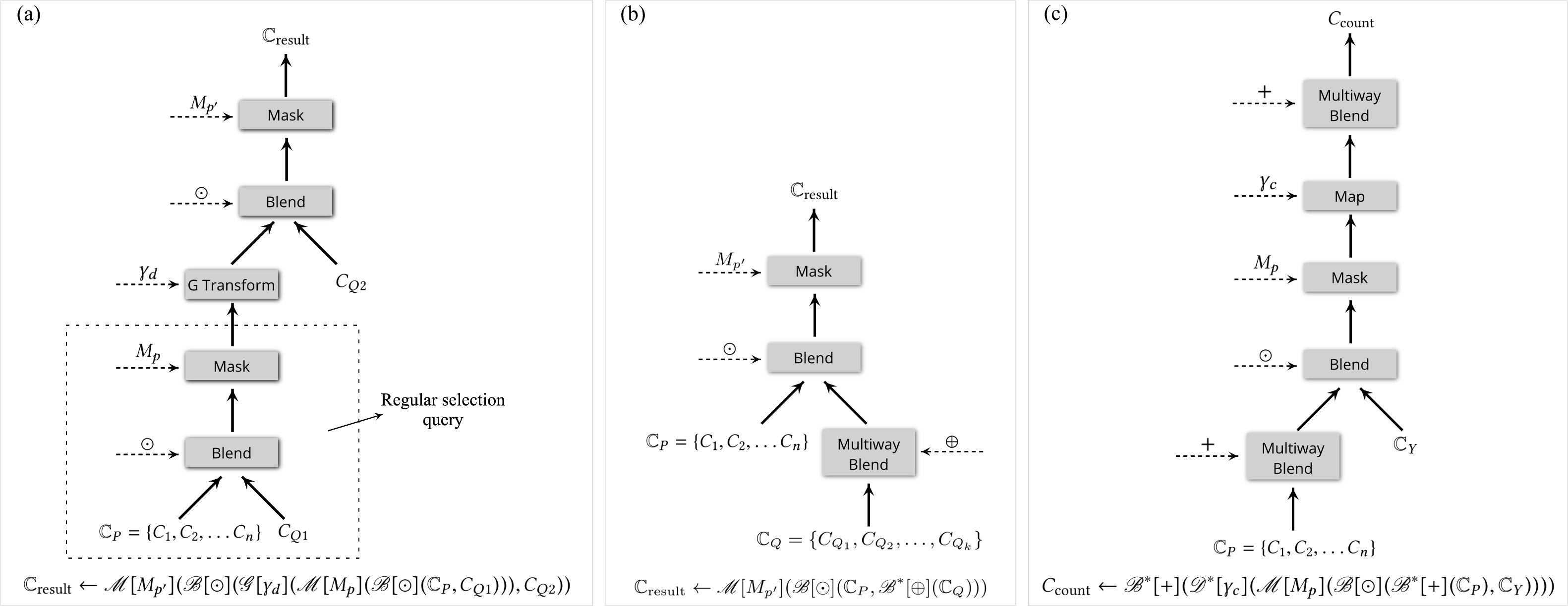}
	\vspace{-0.3in}
	\caption{Examples of alternate plan execution strategies for complex queries.
		(a)~Query plan for a selection query over origin-
		destination data having a polygonal constraint on both spatial attributes.
		(b)~Selection query with multiple polygonal constraints.
		(c)~Spatial aggregation approach used in \protect\cite{rasterjoin}.
	}
	\label{fig:complex-queries}
	\vspace{-0.3cm}
\end{figure}

Let $\C_p$ be the canvases corresponding to $D_P$ defined
as before, but with respect to the origin attribute.
Let $C_{Q1}$ and $C_{Q2}$ be canvases corresponding to the query 
constraints defined as follows:
\begin{eqnarray}
C_{Qi}(x,y)[0] &=& \emptyset \nonumber \\ 
C_{Qi}(x,y)[1] &=& \emptyset \nonumber \\ 
C_{Qi}(x,y)[2] &=&  
\begin{cases}
(i,1,0) & \text{if  } (x,y) \text{ falls inside } Q_i \\ \nonumber
\emptyset & \text{otherwise}
\end{cases} \nonumber
\end{eqnarray}
The above query can then be realized as:
\[
 \C_{\text{result}} \leftarrow \opM[M_{p'}](\opB[\odot](\opG[\gamma_d](\C_{\text{origin}}),C_{Q2}))
\]
\itemspace
where
\[
\C_{\text{origin}} \leftarrow \opM[M_p](\opB[\odot](\C_P,C_{Q1}))
\]
is the same expression as the selection query used earlier and selects
the points that belong to the intersection between $\C_P$ and $C_{Q1}$.
The function $\gamma_d: S^3 \rightarrow \R^2$ is used to 
transform the point from the origin to the destination location 
and is defined as 
\[
\forall s \in S^3, \gamma_d(s) = \textit{\text{destination}}(s[0][0])
\]
Here, \textit{destination()} is a function that takes the $id$ of the point
as input and returns the destination location;
and the mask function $M_{p'}$ defined as:
\[
M_{p'} = \left\lbrace s \in S^3 \mid s[0] \neq \emptyset \text{ and } s[2][0] = 2 \right\rbrace 
\]
The other parameter functions $M_p$ and $\odot$ are defined as before.
Figure~\ref{fig:complex-queries}(a) illustrates the above expression as a plan diagram.
Intuitively, this plan first computes $\C_{origin}$,
  \ie all records whose origin intersect with $Q_1$. It then transforms
  each record in $\C_{origin}$ to its destination and tests for their
  intersection with $Q_2$.

%% file: implementation.tex
\section{Notes on Implementation}
\label{sec:implementation}

We now briefly describe a proof-of-concept GPU-based implementation 
of our model to demonstrate its advantages with respect to enabling the
reuse of operators.
In particular, we discuss the blend and mask operators required to
implement spatial selection queries.
To further illustrate the expressive power of our model, we also examine the
spatial aggregation operation proposed
in~\cite{rasterjoin}, and show how it translates directly into an
algebraic expression.
We note that there can be alternate implementations and that different
design choices can be made, but these are beyond the scope of this
paper.

\newcommand{\includeQueryImage}{}
\newcommand{\includeQueryText}{Figure~\ref{fig:complex-queries}(b) shows the plan for this query.\xspace}

\input{impl-core}

\subsection{Spatial Aggregation}
\label{sec:impl-spatial-aggregation}

Consider the spatial join-aggregation query 
discussed in Section~\ref{sec:aggregate-queries}.
Recall that the typical evaluation strategy used by existing systems
is to perform a join followed by an aggregation.
\textit{\rasterjoin}~\cite{rasterjoin} proposed 
an alternate approach that maps these queries into
operations supported by the graphics pipeline in
GPUs, leading to orders of magnitude 
speedup over CPU-based approaches.
\rasterjoin can be directly mapped into a query execution plan
using the proposed spatial operators as illustrated 
in Figure~\ref{fig:complex-queries}(c), and 
translates to the following expression:
\[
C_{\text{count}} \leftarrow \opB^*[+](\opD^*[\gamma_c](\opM[M_p](\opB[\odot](\opB^*[+](\C_P),\C_Y))))
\]
Here, the parameters $+,\gamma_c,M_p$, and $\odot$ are the same as defined earlier.
While the above expression assumes \textit{count} as the aggregation
function, other aggregations can be incorporated by modifying the 
blend operation parameter $+$ appropriately.
Note that in this plan, all the points are first merged into a single
canvas which keeps track of partial aggregates. 
That is, each canvas pixel maintains the
count of all points that fall into that pixel.
This canvas is then joined
with the set of input polygons to identify the points that
intersect with the polygons, and the results are again merged to
compute the final aggregate. 
In other words, the counts from the individual pixels
that fall within a polygon are combined to generate
the aggregation for that polygon.
Even though this approach performs an additional merge (through the multiway blend), 
the size of the input for the join is drastically reduced 
(there is only one canvas on the left hand side of the blend),
thus reducing the cost of the entire plan.

%% file: impl-core.tex
\subsection{Proof-of-Concept Prototype}
\label{sec:impl-spatial-selection}

The prototype was implemented using C++ and OpenGL.
It assumes the traditional representation of
point and polygon data sets, that is, they are stored as a set of tuples.
Instead of duplicating the geometric objects
in the data by explicitly storing the corresponding 
canvases, we create the canvases on the fly when a query is
executed. 

\myparagraph{Data Representation.}
Recall from Section~\ref{sec:representation} that geometric objects are modeled as a union of smooth manifolds, and a canvas representing these objects
are defined as a scalar function over $\R^2$. Given such a continuous formal representation, it is therefore important to have a discrete representation
to be used in the implementation.
Our choice was to maintain a canvas as a \textit{texture}~\cite{redbook},
which corresponds to a collection of pixels. Here, each pixel
stores the object information triple. 
The canvas functions are defined as
discussed in Section~\ref{sec:select-queries}.
However, since the pixels discretize the space, 
it is also necessary to store additional data corresponding to the  geometry boundaries.
In the case of points, this additional information corresponds to the 
actual location of the points. 
For polygons, we store a flag that is set to true if 
the pixel is on the boundary of the polygon. 
To accurately identify all boundary pixels,
we use an OpenGL extension that enables conservative rasterization. 
This identifies and draws all pixels that are touched by a triangle (or line), 
and is different from the default rasterization, wherein a pixel is only drawn when $>50\%$ of the 
pixel is covered by the primitive. This ensures that the border pixels are kept track of 
in a conservative fashion, and hence there is no loss in accuracy.
Additionally, a simple index is maintained
that maps each boundary pixel to the actual vector 
representation of the polygon.

The canvases are created on the fly by simply
rendering (\ie drawing) the geometry using the traditional
graphics pipeline. The color components (r,g,b,a) are used to store the 
canvas function. This rendering is performed onto an off-screen
buffer, which generates the required texture.
To handle polygons with holes, the outer polygon is first drawn
onto the off-screen buffer. The inner polygon (representing one or more holes)
is then drawn such that the pixels corresponding to it are negated 
(\ie the canvas function is set to null).

\myparagraph{Operators.}
The \emph{blend operator} is accomplished through a straightforward 
\emph{alpha blending}~\cite{redbook} of two textures, which is
supported as part of the graphics pipeline.
The \emph{mask operator} looks up each pixel of the texture in parallel
and tests for the mask condition. Note that here, the 
boundary information is used to perform an accurate test
if the point is part of a pixel that is on the boundary of the
polygon.

Note that if an approximate result suffices, then the hybrid
representation of the canvas can be entirely eliminated, making
the implementation simpler. In fact, in this case,
the texture size can be adjusted in order to appropriately
bound the error in the query result, similar to the approach used
in~\cite{rasterjoin}.

\myparagraph{Alternate Implementations.}
Another possibility for the implementation is to represent geometric objects as 
a collection of simplicial complexes, thus avoiding any rasterization. 
The operators then can be implemented to make use of the 
native ray tracing support provided by the latest RTX-based Nvidia GPUs.
We decided to use the rasterization pipeline instead so that our
prototype could support any modern GPU from multiple vendors, and not 
just the RTX GPUs from Nvidia.

\myparagraph{Queries.}
%
%
The \emph{polygonal selection of points} is 
accomplished by first creating the canvases corresponding 
to the query polygon and query points, which are 
blended together and then filtered using the mask operator.
The operator functions are as defined previously.
Our implementation, without any modification, also works for
\emph{polygonal selection of polygons}, i.e., if the input is changed
from a set of points to a set of polygons.

\includeQueryImage

A straightforward variation of the selection query is to \emph{support
  multiple polygons as part of the constraint}. In particular,
consider the case when the constraint requires the input point to be
inside at least one of the polygons (a disjunction).
Existing approaches accomplish this by testing the points
with respect to each of the polygonal constraints. However,
using our model this query can be expressed as follows 
using just the blend and mask operators:
\[
 \C_{\text{result}} \leftarrow \opM[M_{p'}](\opB[\odot](\C_P,\opB^*[\oplus](\C_Q)))
\]
Here, $\C_Q$ is the collection of canvases corresponding to the
query polygons, and $\odot$ and $\oplus$ are the blend functions
defined in Section~\ref{sec:select-queries}. 
The above expression first blends together all the 
query constraint polygons into a single canvas,
which is then used to perform the select similar
to the single polygon case.
The mask function $M_{p'}$ is defined as:
\[
M_{p'} = \left\lbrace s \in S^3 \mid s[0] \neq \emptyset \text{ and } s[2][0] \geq 1 \right\rbrace 
\]
Recall that the mask function $M_{p}$ used for the single query polygon case
tests the incidence of the polygon on a pixel by testing the $id$
field of the function value corresponding to 2-primitives.
Instead, this is accomplished using $M_{p'}$ by
checking if the count of the polygons incident on the pixel is 
at least one.
Thus, this mask function $M_{p'}$ is valid even when there is only a single query polygon.
So, in our implementation, we use this instead of the $M_{p}$ defined earlier.
\includeQueryText
Furthermore, as we discuss in Section~\ref{sec:evaluation}, using
the proposed operators also helps improve the performance 
of the queries when compared to the traditional approach.
A query with a conjunction can be expressed similarly, by appropriately
adjusting the mask function.

%% file: exp.tex
\section{Experimental Evaluation}
\label{sec:evaluation}

We now briefly discuss the performance of the spatial selection 
queries using the prototype described above. 
All experiments were run on a \emph{laptop} having an Intel Core i7-8750H
processor, 16~GB memory and 512~GB SSD. The laptop has a dual Nvidia
GTX~1070 Max-Q GPU with 8~GB graphics memory, and an integrated Intel
UHD Graphics 630 GPU.

\myparagraph{Data and Queries.}
The main goal of our evaluation is to: 
1)~demonstrate the advantage of using GPU-friendly operators
compared to a traditional GPU-based solution; and
2)~illustrate how the same operators can be used for variations
of a given query.
We do this using selection queries that selects 
trips from the New York City's taxi data having their
pickup location within a query polygon. 
To demonstrate point (2), we also
use queries having a disjunction of multiple polygonal
constraints. 
The size of the input is varied using the pickup time range of the taxi trips.

To mimic real-world use cases, all the query polygons used 
in these queries were ``hand-drawn" using a 
visual interface (\eg~\cite{taxivis}) and adjusted to have the same 
MBR. 
We then use as input only taxi trips that
have their pickup location within this MBR.
In other words, 
the evaluation
assumes the existence of a filtering stage and 
primarily focuses on the refinement step.
We decided to do this for two reasons.  First, the
refinement stage, and not filtering, is now the primary
bottleneck. Unlike previous decades 
when the disk-based index filtering was the primary bottleneck,
due to the existence of fast ssd-based
storage and large CPU memory, the filtering takes only a small
fraction of the query time. For example, the filtering step 
used by the state-of-the-art GPU-based selection approach, 
even though it is CPU-based, 
takes only a few milliseconds even for data having over a billion points~\cite{STIG2016}.
Second, when working with complex queries, depending on the query parameters, 
the optimizer need not always choose to use the spatial index corresponding to a 
spatial parameter, and the spatial operations could be further up in the plan
(e.g., the optimizer might to choose first filter based on another attribute, say time, before 
performing a spatial operation).
In such scenarios, the spatial operation would not have the benefit of an index based filtering,
and query bottleneck would then be the refinement step.
Additionally, the above setup also helps remove input bias when comparing the
performance across polygonal constraints having
different shapes and sizes.

\myparagraph{Approaches.}
We compare the performance of our approach with a CPU baseline, 
a parallel CPU implementation 
using OpenMP, as well as a GPU baseline.
Because of the above mentioned experimental setup
that eliminates the effect of indexes used by
current state-of-the-art,
we only need to implement the PIP tests
for the above baselines.
While our approach was executed on two different GPUs (denoted as Nvidia and Intel), 
the GPU baseline was executed only on the faster Nvidia GPU.

\begin{figure}[t]
	\centering
	\includegraphics[width=\linewidth]{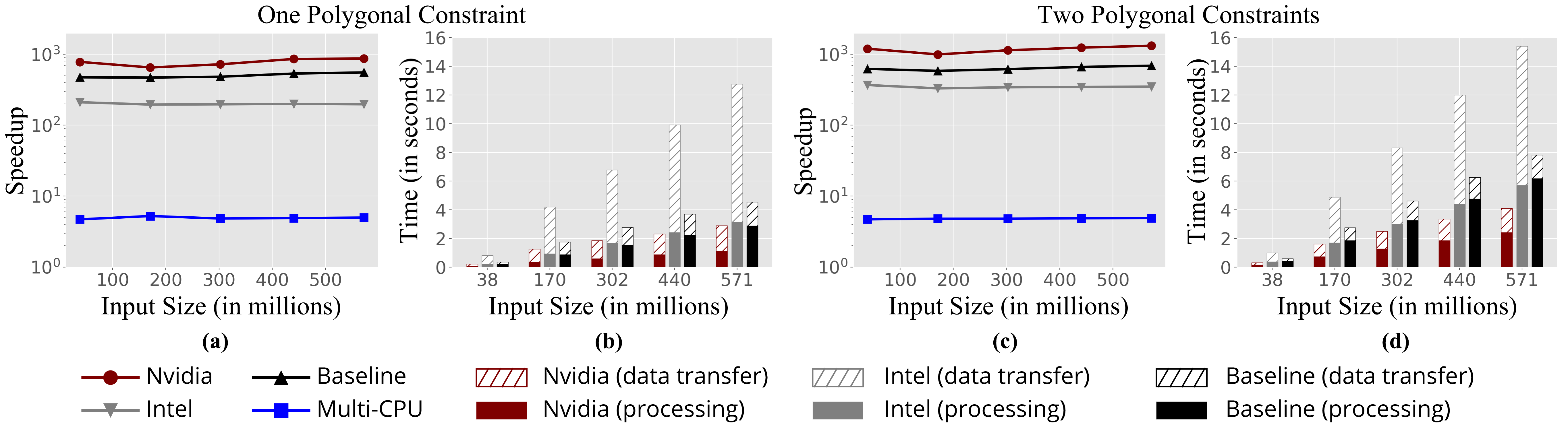}
	\vspace{-0.2in}
	\caption{Scaling with input size.}
	\label{plot:scaling}
	\vspace{-0.1in}
\end{figure}

\myparagraph{Performance.}
Figures~\ref{plot:scaling}(a) and \ref{plot:scaling}(c) shows the speedup achieved by 
the different approaches over a single threaded CPU implementation
when the query had one and two polygonal constraints, respectively. 
Note that while all GPU-based approaches are over two orders of
magnitude faster than the CPU-based approach, the speedup
of our approach increases when the number of polygonal constraint increases.
This is because, the only additional work done by our approach 
when there are additional polygons is to blend the constraint polygons. 
This is significantly less work when compared to existing approaches 
which have to perform more PIP tests in this case.
This is corroborated when looking at the query run times in
Figures~\ref{plot:scaling}(b) and \ref{plot:scaling}(d) 
wherein our approach (in red) requires only 4 seconds (using the Nvidia GPU)
even when there are two polygons as constraints even when the query MBR
has as many as 571M points.
For a given input and GPU, not only is the time to transfer data between the CPU and GPU
similar, but is also a significant fraction
of the query time. In this light, the speedup in the processing time 
achieved using our model over a traditional GPU-based approach (which is greater than 
the overall speedup depicted in Figures~\ref{plot:scaling}(a) \& (c)) clearly
demonstrates the advantages of using a GPU-friendly approach.

\begin{figure}[h]
	\centering
	\includegraphics[width=0.58\linewidth]{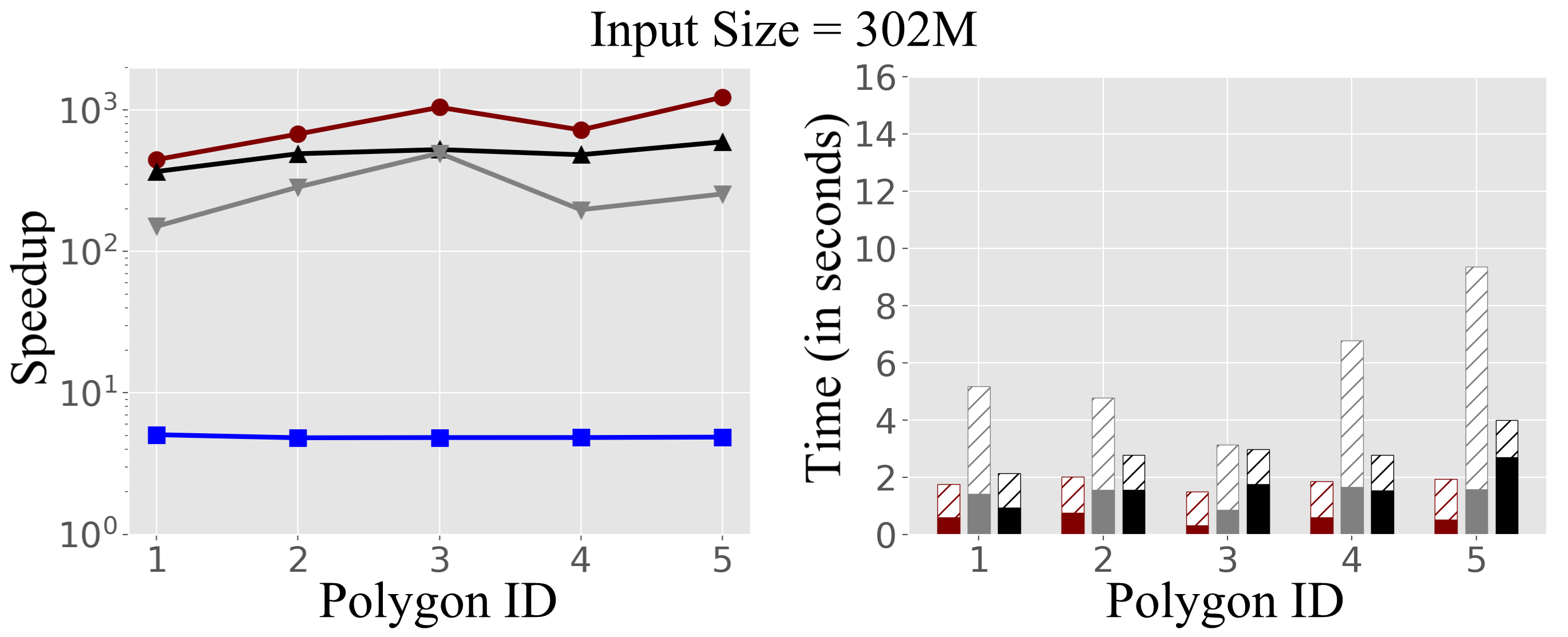}
	\vspace{-0.1in}
	\caption{Varying polygonal constraints. Legend is the same as in Figure~\ref{plot:scaling}.}
	\label{plot:polygon-scaling}
	\vspace{-0.1cm}
\end{figure}

Figure~\ref{plot:polygon-scaling} shows the speedup and running times
when the polygonal constraint is varied. The different
polygons had different shapes (and sizes) with query selectivity varying from 
roughly 3\% to 83\%. 
While there is some variation in the processing time depending on the
complexity of the polygon constraint, this variation is
higher for the baseline. This is because the number of PIP
tests performed by the baseline is linearly proportional to the size
of the polygon.
Irrespective of this complexity,
our approach using the discrete GPU requires at most 2 seconds
even when the MBR has 302M points.

Also interesting to note is the performance of our approach 
on the integrated Intel GPU. 
While, as expected, it is slower than the GPU baseline 
using a Nvidia GPU, it is still over 2-orders of magnitude 
faster than the CPU implementation. 
Given that these GPUs are
present in even mid-range laptops, ultrabooks, and even tablets,
our algebraic formulation
can potentially allow fast spatial queries even on such
systems.

%% file: discussion.tex
\section{Discussion}
\label{sec:discussion}

\itemspace
\myparagraph{Interoperability with Relational Model.}
The proposed model is compatible with
the relational model and can be incorporated
into existing relational systems. 
Recall that the minimalistic definition of $S$ used in Section~\ref{sec:canvas}
reserves the first element of the triple to store 
the unique ID corresponding to the data record. Thus,
given a set of canvases corresponding to existing data sets,
it is possible to switch to the corresponding relational tuple 
using this ID. 
Analogously, the storage structure of a relational tuple can 
be changed to link to the corresponding canvas, thus allowing
connection in the opposite direction. Alternatively,
similar to our proof-of-concept implementation, the
canvases could also be created on demand.

Thus, conceptually, one can consider the relational tuple and a canvas
to be the \textit{dual} of each other allowing a seamless use of the
two representations by a query optimizer to appropriately 
generate query plans involving both spatial and non-spatial operators.

\myparagraph{Query Optimization.}
The proposed model facilitates query optimization in the following ways.

\myparagraphem{1. Allowing different query execution plans.}
Given a complex query $Q$, the proposed model enables the creation
of multiple plans to realize $Q$.
We gave two such examples in the previous section---one for
disjunction and the other for spatial aggregation.
In all such scenarios, by appropriately modeling the cost functions
of the operators together with metadata about the input,
the optimizer can choose a plan that has a lower cost.

\myparagraphem{2. Supporting diverse implementations.}
It is also possible to have multiple implementations
of the same operators, for example, using pre-built
spatial indexes. Each of the indexes would
result in a different cost based on the properties 
of the data and the query,
thus providing a rich set of options over which to
perform the optimization.

\myparagraphem{3. Enabling general query processing.} 
Given the duality between the canvas and the relational tuple as discussed above,
the proposed operators can also be easily plugged into existing
query optimizers, thus allowing for complex queries
involving both the spatial and relational attributes.

\myparagraph{Limitations.}
While the proposed data representation
can be directly extended to support 3D primitives, the proposed operators
over such 3D data do not have a straightforward implementation using the GPU.
Given that native ray tracing support is now being introduced in GPUs,
it would be interesting to explore extensions to our algebra that
make use of such advances to support 3D spatial queries.

%% file: rel-work.tex
\section{Related Work}
\label{sec:rel-work}

\itemspace
\myparagraph{Spatial Queries.}
The most common approach used for 
executing spatial queries is to implement custom techniques for
different query types.
Selection queries, for example, are typically handled through the 
use of spatial indexes. These include R-Trees~\cite{Guttman84}, R$^*$-trees~\cite{Beckmann90},
kd-trees~\cite{Bentley75}, quad trees~\cite{quadtree} and the grid index~\cite{gis-book}.
While such indexes are also useful for other query types such as
spatial joins, enhancements and more efficient algorithms are often
designed for specific queries.
For instance, several works focus on the filtering step of 
spatial join algorithms~\cite{spatialjoinsurvey, rtreejoin, pbsm96, transformers}. 
Custom algorithms have also been designed for spatial
aggregation~\cite{sptempaggrsurv,aggregateprocessing,onlinespatialaggr,aRtree}
and for nearest neighbor-based queries (see
e.g.,~\cite{Jagadish05,Yianilos93,Katayama1997,Hjaltason1999,Roussopoulos1995}).

The advent of modern hardware with multiple processing units has
led to the design of new approaches that use them for spatial query
processing. In particular, GPUs, and
clusters supporting the MapReduce paradigm, are extremely popular for this
purpose.
GPUs have been used for spatial selections~\cite{STIG2016},
spatial joins~\cite{Zhang2015,gcmf}, spatial aggregations~\cite{rasterjoin},
as well as nearest neighbor queries~\cite{Bustos2006,Pan2011}. 
Similarly, there are dedicated spatial database systems designed using MapReduce
such as Hadoop-GIS~\cite{hadoop-gis} and Simba~\cite{Xie:2016:SEI:2882903.2915237}.
Eldawy and Mokbel~\cite{Eldawy2016} provide a comprehensive survey of
approaches that use MapReduce for spatial query processing.
Covering the numerous work related to different spatial queries
is beyond the scope of this paper. However, note that
many of these approaches (\eg indexes) can be easily applied to
enhance the operators of the proposed algebra.

\vspace{-0.1cm}
\myparagraph{Spatial Data Models and Algebras.}
\input{spatial-algebra-related}

%% file: spatial-algebra-related.tex
Specific to spatial databases, 
G{\"u}ting~\cite{Guting88} introduced geo-relational algebra,
which extends relational algebra to
include geometric data types and operators. 
The geometric data types included points, lines, and polygons (without
holes) and the geometric operators included operations that are now common 
in most spatial database solutions (containment, intersection, perimeter, area, etc.). 
Aref and Samet~\cite{Aref1991a,Aref1991b,Samet1995} generalized the above model
and provided one of the first high-level discussions
on integrating spatial and non-spatial data to build a spatial
database system, and the related challenges involved in
designing a query optimizer for such a system.
Note that current spatial extensions 
follow approaches very similar to the ideas proposed in these works.
This model is \emph{user facing}, \ie
the queries of interest are expressed making use of the
data types and the operators provided in the model. The implementation
of the operators, however, is left to the developer, and
often devolves into having separate implementations for each data type/query combination 
(similar to the selection query example discussed in Section~\ref{sec:intro}).
Our approach, on the other hand, which uses a single representation for all spatial data types
and set of GPU-friendly operators different from the traditional operators,
is \emph{primarily meant to help database developers implement
an efficient GPU-based spatial query engine}: it can be 
incorporated into existing systems unbeknownst to the user
while at the same time providing significant benefits to
the database engine and query performance.

Different from the extended relational models, Egenhofer and Franzosa~\cite{EGENHOFER91}
proposed a model that uses concepts from point set topology
for spatial queries. In particular, this work models spatial data objects (of a single type, like lines or regions)
as closed sets (that defines the underlying topological space), and uses the topological relationship between pairs of closed sets
to answer spatial queries. These relationships are computed based on
9 possible intersections computed between the open set, boundary and complement corresponding to the closed sets.
Egenhofer and Sharma~\cite{Egenhofer93} showed the equivalence of the above model to a raster space, 
thus making it suitable for GIS queries involving raster data.
Kainz~et~al.~\cite{Kainz93} model the same topological relations as above, but using partially ordered sets.
While theoretically elegant, there are three main shortcomings of this
topological approach: (1)~the topological relationships are tied to a
particular data type, that is, between two regions, or two lines,
etc., making it difficult to work with complex spatial objects;
(2)~computing the relationships requires costly intersection tests to
be performed between every pair of spatial objects, making the
approach untenable for working with large spatial data sets; and more
importantly, (3)~while intersection-based queries are straightforward,
distance-based queries such as distance-based selections/joins,
nearest neighbors etc. cannot be expressed using this model.

Gargano~et~al.~\cite{GARGANO1991} proposed an alternative model
that supports complex objects in which spatial objects are represented
using a set of rectangular regions.  The spatial queries are then
realized as operations over these sets.  This representation results
in a loss of accuracy in the query results.  Trying to overcome this
using very small rectangles can result in high memory overheads and
also require expensive set operations, thus limiting the practical
applicability of this model.

G{\"u}ting and Hartmut proposed another model called
Realms~\cite{Guting93} and a corresponding ROSE
algebra~\cite{Guting1995}.  A Realm models spatial data as a planar
graph, where the nodes correspond to points on an integer grid. Given
the hardware limitations during that era, the goal of this approach
was to avoid costly floating point operations and any imprecision in
the query computation. As data is inserted into the database, the
spatial objects are ``redrawn" to ensure topological consistency (such
as locations of intersection points).
This framework has important limitations.  First, even
though the redrawings ensure that queries involving
intersection tests can be efficiently and precisely computed using
only integer operations, due to the distortion involved,
distance-based queries now become imprecise.
Second, it is necessary for all query parameters to be a part of the Realm.
Thus, when generating dynamic queries (common in several data analysis tasks), 
the query parameters
have to first be inserted into the Realm, 
requiring several redrawings of the existing data.
Then, after query execution, the newly inserted parameters should be removed. 
Not only is this expensive, the redrawings caused by the temporary insertions
are also not undone, resulting in further distortions.
Third, queries involving spatial objects outside the Realm boundaries
are not possible. This is a major drawback in modern exploratory data
analysis tasks where users can dynamically change their focus
as they test and formulate new hypotheses. 
Finally, similar to the extended relational models, there are separate
data types for points, lines, and polygons, making the implementation
specific to these data types, and also making it difficult to
incorporate complex spatial objects consisting of more than one type.

All of the above models/algebras were designed before GPUs
became mainstream, and thus an implementation of these models using
GPUs is non-trivial (difficult to parallelize, involves iterative
algorithms like intersection computations, etc.).
Our approach on the other hand was designed keeping GPUs in mind,
and is based on computer graphics operations for which they are optimized.

Models have also been proposed that focus on moving
objects~\cite{Nidzwetzki2017} which are orthogonal to our work.  For
GIS applications, Tomlin~\cite{tomlin1994map} proposed the Map algebra
which was then extended to support time by Jeremy et
al.~\cite{Jeremy05}.
The map algebra was designed to enable cartographers to easily specify
common cartographic functions.
Voisard and David~\cite{Voisard2002} propose a layered model specific to geographic maps
to help users build new maps.
From an implementation point of view, all of the above operations can be translated into spatial queries
for execution, and thus an efficient spatial model will be useful in such scenarios as well.

%% file: conclusion.tex
\vspace{-0.1in}
\section{Conclusion}
\label{sec:conclusion}

In this paper, we introduced a GPU-friendly data model 
and algebra to support queries
over spatial data sets. A key and novel idea in this work
is to use a representation that captures the geometric properties
inherent in spatial data, and design GPU-friendly operators that can be applied
directly on the geometry. We have shown that the proposed algebra is
expressive and able to realize common spatial queries. In addition,
since the algebra is closed, it can also be used to construct complex
queries by composing the operators.
The potential ease of implementation afforded by our approach, 
and its performance even on commodity hardware,
can greatly influence the design as well as adoption of GPU-based
spatial database solutions, thus democratizing real-time spatial
analyses. This is corroborated by the
results of the experimental evaluation carried out with our
proof-of-concept prototype.

We also believe that this work opens opportunities for new research
in the area of spatial query optimization, both for the development of
new theory, including algorithms for plan generation and cost
estimation, and systems that use the algebra to efficiently evaluate
queries over the growing volumes of spatial data.